\documentclass[aps,prd,groupedaddress,showpacs,graphicx,nofootinbib]{revtex4}
\usepackage{amssymb,graphics,graphicx,color}
\def\lae{\;^{<}_{\sim} \;} \def\gae{\; ^{>}_{\sim} \;}
\begin{document}

\title{Kinematically Blocked Curvaton}
\author{Lu-Yun Lee$^1$}\email{d9522809@oz.nthu.edu.tw}
\author{Chia-Min Lin$^{1,3}$}\email{cmlin@phys.nthu.edu.tw}
\author{Chian-Shu Chen$^{2,3}$}\email{chianshu@phys.sinica.edu.tw}
\affiliation{$^1$Department of Physics, National Tsing Hua University, Hsinchu, Taiwan 300}
\affiliation{$^2$Physics Division, National Center for Theoretical Sciences, Hsinchu, Taiwan 300}
\affiliation{$^3$Institute of Physics, Academia Sinica, Taipei, Taiwan 115}

\date{Draft \today}

\begin{abstract}
In this paper, we investigate the idea that the decay of a curvaton is kinematically blocked and show that the coupling constant for curvaton decay can be as large as ${\cal O}(1)$.
We also find in this case the lower bound of the Hubble parameter at horizon exit from big bang nucleosynthesis (BBN) is $H_\ast \gae 7.2 \times 10^{-9}M_P \sim 10^{10} \mbox{ GeV}$. Similar to conventional curvaton scenario, the nonlinear parameter can be as large as $f_{NL}=100$.
\end{abstract}
\maketitle

\section{Introduction}
The primordial density (curvature) perturbation \cite{Lyth:2009zz, Mazumdar:2010sa} provides the seeds of large scale structure formation and the Cosmic Microwave Background (CMB) temperature fluctuation. The CMB tells us that the spectrum of curvature perturbation is $P_\zeta^{1/2} \sim \zeta =5 \times 10^{-5}$ and we call this \emph{CMB normalization}. An elegant method to calculate the curvature perturbation $\zeta$ is the \emph{delta N formalism} \cite{Sasaki:1995aw, Sasaki:1998ug, Lyth:2004gb, Lyth:2005fi} which states:
\begin{equation}
\zeta=\delta N=N_\phi \delta \phi + \frac{1}{2} N_{\phi\phi} (\delta \phi)^2 + \cdots,
\label{eq0}
\end{equation}
where $N$ is number of e-folds and $\delta \phi \sim H/2\pi$. Here $\phi$ may be the inflaton field $\psi$ or some other field (like a curvaton $\sigma$) which is light during inflation and affects $N$. The subscript $_\phi$ means derivative with respect to $\phi$. Note that $\delta \phi$ is calculated on a flat slice (gauge) and $\zeta$ is defined to be the curvature perturbation on a uniform density slice. Therefore $\delta N$ is the difference of $N$ between these two slices. It is convenient to parameterize the second order effect by a nonlinear parameter $f_{NL}$ defined by
\begin{equation}
\zeta=\zeta_g+\frac{3}{5}f_{NL}\zeta^2_g+\cdots,
\end{equation}
where $\zeta_g$ denotes the Gaussian (first order) part of $\zeta$. From Eq.~(\ref{eq0}) we can see that\footnote{In this paper, we only consider \emph{local type} non-Gaussianity.}
\begin{equation}
f_{NL}=\frac{5}{6}\frac{N_{\phi\phi}}{(N_\phi)^2}.
\label{fnl}
\end{equation}
Currently the upper bound of $f_{NL}$ is roughly given by ($2$-$\sigma$) \cite{Komatsu:2010fb}
\begin{equation}
|f_{NL}| \lae 100.
\label{ex}
\end{equation}
In the near future, the PLANCK satellite will reduce the bound to $|f_{NL}|<5$ if non-Gaussainity is not detected.

The number of e-folds is related to the inflaton field $\psi$ via
\begin{equation}
N=\frac{1}{M_P^2} \int^\psi_{\psi_{end}} \frac{V}{V^\prime}d\psi
\end{equation}
 The $\zeta$ produced by the quantum fluctuation of the inflaton field in the simplest single-field slow-roll inflation models is given by
\begin{equation}
\zeta=\delta N = \frac{1}{M_P^2} \frac{V}{V^\prime}\delta\psi \sim \frac{H_\ast}{2 \pi \sqrt{2\epsilon}},
\end{equation}
where
\begin{equation}
\epsilon \equiv \frac{M_P^2}{2} \left( \frac{V^\prime}{V} \right)^2
\end{equation}
is a slow-roll parameter and the subscript $_\ast$ denotes horizon exit.
After imposing CMB normalization and adopt a typical value of $\epsilon \lae 0.01$, we have
\begin{equation}
H_\ast \lae 10^{-5}M_P.
\label{eq8}
\end{equation}
The inequality is saturated if the curvature perturbation is dominated by the inflaton field.
Single-field slow-roll inflation would be ruled out if large non-Gaussianity is detected \cite{Maldacena:2002vr}.

It is well known that $\zeta$ can also come from a curvaton field $\sigma$ \cite{Lyth:2001nq, Enqvist:2001zp, Moroi:2001ct}. The energy density of the curvaton $\rho_\sigma$ is by definition subdominant during inflation. After inflation, the inflaton decays and the universe is dominated by radiation. However, the curvaton is assumed to decay later. As the universe expands and the Hubble parameter decreases to $H \sim m_\sigma$, the curvaton field would start to oscillate with an energy density $\rho_{\sigma,o}=\frac{1}{2}m^2_\sigma \sigma_\ast^2$. At this time the energy density of the universe is dominated by radiation and is given by $\rho_{tot,o}=3 m^2_{\sigma} M_P^2$. Eventually the curvaton decays with an energy density $\rho_{\sigma,D}$. The total energy density is denoted as $\rho_{tot,D}$ when curvaton decays. Since radiation dilutes as $\sim a^{-4}$ and matter dilutes as $\sim a^{-3}$, we have
\begin{equation}
\rho_{\sigma,D}=\rho_{\sigma_o}\left(\frac{\rho_{tot,D}-\rho_{\sigma,D}}{\rho_{tot,o}}\right)^{3/4}
\label{eq1}
\end{equation}
The number of e-folds is related to the curvaton field $\sigma$ via
\begin{equation}
N=\frac{1}{3}\ln \left(\frac{\rho_{\sigma,o}}{\rho_{\sigma,D}}\right).
\end{equation}
Therefore by using Eq.~(\ref{eq0}), we can obtain (to first order)
\begin{equation}
P^{1/2}_\zeta = N_\sigma \delta \sigma= \frac{1}{3 \pi} \Omega_{\sigma,D} \frac{H_\ast}{\sigma_\ast},
\label{eq4}
\end{equation}
where
\begin{equation}
\Omega_{\sigma,D} \equiv \frac{3 \rho_{\sigma,D}}{4(\rho_{tot,D}-\rho_{\sigma,D})+3\rho_{\sigma,D}}
\label{eq3}
\end{equation}
roughly represents the energy ratio of the curvaton at decay compared with the total energy density of the universe.
From Eq.~(\ref{fnl}), we can obtain the nonlinear parameter
\begin{equation}
f_{NL}=\frac{5}{6}\frac{N_{\sigma\sigma}}{N^2_\sigma}=\frac{5}{4}\frac{1}{\Omega_{\sigma,D}}-\frac{5}{3}-\frac{5}{6}\Omega_{\sigma,D}.
\label{eq13}
\end{equation}

Let us assume the decay rate of the curvaton is given by
\begin{equation}
\Gamma=\frac{g^2}{8\pi}m_\sigma.
\end{equation}
Conventionally, the curvaton decays when $H \sim \Gamma$ with $\rho_{tot,D}=3 \Gamma^2 M_P^2$. In order to obtain a significant energy density of the curvaton when it decays, the lifetime of the curvaton should be long enough which makes $g<10^{-6}$ (see, for example \cite{Lin:2009fk}). This kind of small coupling usually is not favored in particle physics. In this paper, we investigate the cosmological consequences when the decay of the curvaton is kinematically blocked. In particular we will show that the curvaton can still work even when $g \sim {\cal O}(1)$.

This paper is organized as follows. In section \ref{sec2}, we derive the relevant equations for a kinematically blocked curvaton and explore the parameter space. Section \ref{conclu} is our conclusion.

\section{Kinematically Blocked Curvaton}
\label{sec2}
We assume the curvaton field $\sigma$ coupled to a fermion\footnote{For simplicity, we do not consider the curvaton coupled to other scalar fields $\psi$ through terns like $\lambda^2 \sigma^2 \psi^2$ to avoid complicated issues like preheating or parametric resonance. Even if this kind of coupling exists, it is not guaranteed that parametric resonance can happen. However, this could be an interesting topic for our future work.} field $\chi$ through $g\sigma\overline{\chi}\chi$. The curvaton field starts oscillation when $m_\sigma \sim H$. During the curvaton oscillation, the amplitude $\langle \sigma \rangle \sim \sigma$ gives an effective mass $m_\chi \sim g \sigma$ to the $\chi$ field. Therefore even when $\Gamma \sim H$, the curvaton may still be kinematically blocked and cannot decay. The oscillation amplitude $\sigma$ decreases with the expansion of the universe and the curvaton can decay only when $\sigma_D \sim m_\sigma/g$ is achieved. This fact is well known in the context of reheating and was first applied to curvaton in \cite{Enqvist:2010ky}. When the curvaton decays, the decay width is given by
\begin{equation}
\Gamma_D  =  \frac{g^2}{8 \pi}m_\sigma.
\end{equation}
The curvaton energy density at this moment is given by $\rho_{\sigma,D}=m_\sigma^4 / 2g^2$.
By using Eqs.~(\ref{eq1}) and (\ref{eq3}), we can obtain
\begin{equation}
\Omega_{\sigma,D}=\frac{1}{8 \frac{M_P^2}{\sigma^2_\ast}\left(\frac{m^2_\sigma}{\sigma^2_\ast g^2}\right)^{1/3}+1}.
\end{equation}
Therefore from Eq.~(\ref{eq4}), the spectrum is given by
\begin{equation}
P^{1/2}_\zeta=\frac{H_\ast}{3 \pi \sigma_\ast}\frac{1}{8 \frac{M_P^2}{\sigma^2_\ast}\left(\frac{m^2_\sigma}{\sigma^2_\ast g^2}\right)^{1/3}+1}
\label{eq17}
\end{equation}

\begin{figure}[t]
  \centering
\includegraphics[width=0.6\textwidth]{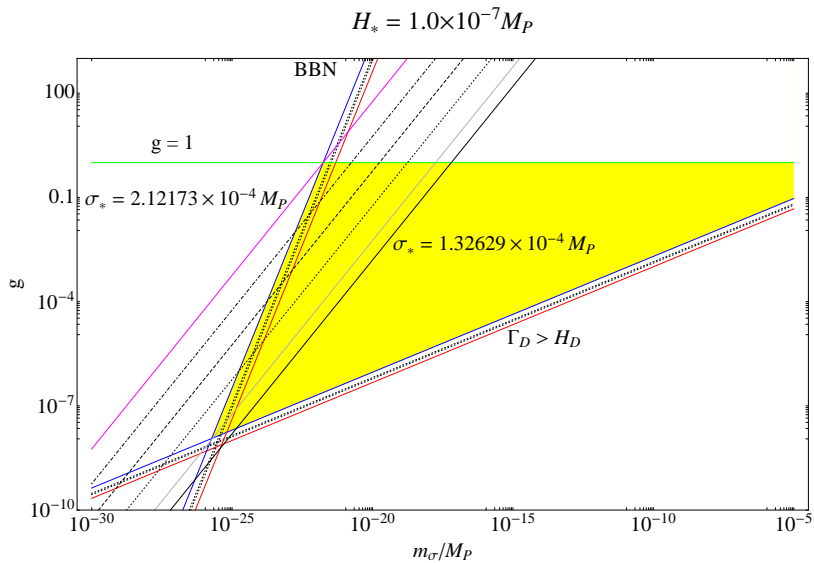}
  \caption{Different $\sigma_\ast$ would corresponds to different lines in the plot. When the lines cross the yellow region, every point on the figure is a viable solution with all the parematers determined. Here we plot several lines by decreasing the field value. The lines go down to smaller $g$ from $\sigma_\ast=2.12173 \times 10^{-4}M_P$ to $\sigma_\ast=1.32629 \times 10^{-4}M_P$ .}
  \label{fig1}
\end{figure}

\begin{figure}[t]
  \centering
\includegraphics[width=0.6\textwidth]{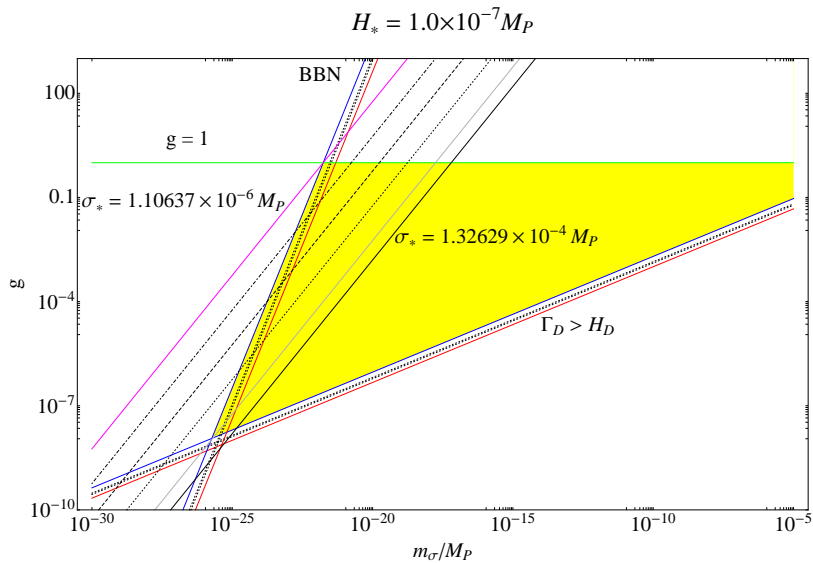}
  \caption{When we decrease the field value further from $\sigma_\ast=1.32629 \times 10^{-4}M_P$ to $\sigma_\ast=1.10637 \times 10^{-6}M_P$, the lines go up to larger $g$ again. If we choose $H_\ast< 10^{-9}M_P$, the lines can never go down and enter the yellow region before they go up again. This means the lower bound of the Hubble parameter is $H_\ast \sim 10^{-9}M_P$.}
  \label{fig2}
\end{figure}

From Eq.~(\ref{eq3}), we can also solve $\rho_{tot,D}$ to obtain
\begin{equation}
\rho_{tot,D}=\frac{m_\sigma^4}{g^2}\left(\frac{3}{8\Omega_{\sigma,D}}+\frac{1}{8}\right).
\end{equation}
There are three constraints in order. First,
in order for a curvaton not to disturb BBN, we require $\rho_{tot,D}>(\mbox{MeV})^4 \sim 10^{-84}M_P^4$. Second, in order to have curvaton starts to decay at $\sigma_D$, we have to make sure $\Gamma_D>H_D$, which implies $3\Gamma^2_D M_P^2 > 3 H_D^2 M_P^2=\rho_{tot,D}$, namely
\begin{equation}
\frac{3g^4}{64 \pi^2}M_P^2 > \frac{m^2_\sigma}{g^2}\left(\frac{3}{8\Omega_{\sigma,D}}+\frac{1}{8}\right)
\end{equation}
Third, we require $g \lae 1$.
We plot all the constraints on FIGs.~\ref{fig1} and \ref{fig2} on a $g$-$m_\sigma$ plane. The upper bound of the mass is given by Eq.~(\ref{eq8}) and the slow-roll condition $m_\sigma < H_\ast$. In the figures, the yellow region is the allowed parameter space. Note that the yellow region does \emph{not} depends on $H_\ast$ (as long as $H_\ast \lae 10^{-5}M_P$), but it depends mildly\footnote{This is the reason why there are two lines for each of the constraint in the figure, one corresponds to $f_{NL}=100$, the other corresponds to $f_{NL}=-5/4$} on $\Omega_{\sigma,D}$ (or equivalently, $f_{NL}$). We choose $f_{NL}=100$ (blue lines) as the boundaries in the yellow region for both BBN and curvaton decay. As an example, we plot some values of $\sigma_\ast$ by taking a typical value of $H_\ast \sim 10^{-7}M_P$. According to the slow-roll condition, here we require $m_g<H_\ast=10^{-7}M_P$. 
By imposing CMB normalization on Eq.~(\ref{eq17}), we find
\begin{equation}
\Omega_{\sigma,D}=\frac{3 \pi \sigma_\ast}{20000 H_\ast}.
\label{omg}
\end{equation}
This means once we fix $H_\ast$, the field value would determine $f_{NL}$ through Eq.~(\ref{eq13}). We list the field values for FIG.~\ref{fig1} (FIG.~\ref{fig2}) with the corresponding $f_{NL}$ in the TABLE~\ref{table1} (TABLE~\ref{table2}).

By imposing CMB normalization in Eq.~(\ref{eq17}), we can write
\begin{equation}
g=
16\sqrt{2}\frac{M_P^3 m_{\sigma}}{\sigma_{*}^4}
\frac{1}{\left(\frac{20000}{3\pi}\frac{H_{*}}{\sigma_{*}}-1\right)^{3/2}}.
\label{eq20}
\end{equation}
The terms inside the parenthesis of the above equation should be positive, otherwise we would have $\Omega_{\sigma,D}>1$ which is impossible. Therefore the viable region of the field value lies in $H_\ast < \sigma_\ast < (20000/3\pi) H_\ast$.

It is clear from Eq.~(\ref{eq20}) that after fixing $H_\ast$, each field value $\sigma_\ast$ corresponds to a line in the $g$-$m_\sigma$ plane. The lines would go down to the lower $g$ region when we decrease $\sigma_\ast$ until some critical value and then the lines would go up when we decrease $\sigma_\ast$ further towards $H_\ast$. This intriguing behavior shows that when we fix $g$ and $m_\sigma$ in the allowed (yellow) region, generically we can have two viable solutions of $\sigma_\ast$ corresponds to different $f_{NL}$ (hence different $\Omega_{\sigma,D}$).
The critical field value $\sigma_\ast^c$ can be obtained by requiring $dg/d\sigma_\ast=0$ and we obtain
\begin{equation}
\sigma_\ast^c=\frac{12500}{3 \pi}H_\ast
\label{eq21}
\end{equation}
It may be interesting to note from Eq.~(\ref{omg}) that no matter what the value of $H_\ast$ is, $\sigma_\ast^c$ always correspond to $f_{NL}=-0.1875$. 
For a low enough $H_\ast$, the line corresponds to the critical field value may not be able to cross the yellow region and we do not have any solution at all. This implies a lower bound of $H_\ast$ in our model. This lower bound can be found by making the critical field value just cross the yellow region, that is $g=1$ and $m_\sigma \sim 1.7 \times 10^{-22}M_P$ (the bound from BBN). Applying these parameters into Eq.~(\ref{eq21}) and imposing CMB normalization, we obtain the lower bound for the Hubble parameter
\begin{equation}
H_\ast \gae 7.2 \times 10^{-9} M_P \sim 10^{10}\mbox{ GeV}. 
\end{equation}

\begin{center}
\begin{table}[h]
\begin{tabular}{|c||c|c|c|c|c|c|} \hline
           $\frac{\sigma_{*}}{M_{P}}$ &$1.32629\times10^{-4}$&$1.94436\times10^{-4}$&$2.08807\times10^{-4}$&$2.1149\times10^{-4}$&$2.12053\times10^{-4}$&$2.12713\times10^{-4}$ \\ \hline 
	  $f_{NL}$ &$-0.1875$&$-1.06597$&$-1.2163$&$-1.24295$&$-1.24849$&$-1.24968$ \\ \hline
\end{tabular}
\caption{\label{table1} The field values and the corresponding $f_{NL}$ for the six lines in FIG.~\ref{fig1}. When the field values are large, the curvaton energy density would be able to dominate the universe when decays, which is the reason for the negative nonlinear parameters.}
\end{table}
\end{center}

\begin{center}
\begin{table}[h]
\begin{tabular}{|c||c|c|c|c|c|c|} \hline
           $\frac{\sigma_{*}}{M_{P}}$ &$1.10637\times10^{-6}$&$2.79248\times10^{-6}$&$7.10247\times10^{-6}$&$1.8461\times10^{-5}$&$5.19677\times10^{-5}$&$1.32629\times10^{-4}$ \\ \hline
	  $f_{NL}$  &$238.084$&$93.3125$&$35.6528$&$12.6294$&$3.23355$&$-0.1875$ \\ \hline
\end{tabular}
\caption{\label{table2} The field values and the corresponding $f_{NL}$ for the six lines in FIG.~\ref{fig2}. Of course the one with $f_{NL}=238$ has already been rule out by Eq.~(\ref{ex}).}
\end{table}
\end{center}

\section{Conclusion}
We have investigated the parameter space of a kinematically blocked curvaton with a quadratic potential. We found a lower bound for the Hubble parameter $H_\ast \gae 10^{10}\mbox{ GeV}$ for this model which is larger than the bound $H_\ast \gae 10^{7}\mbox{ GeV}$ of conventional curvaton scenario \cite{Lyth:2003dt}. Similar to conventional curvaton scenario, a wide range of the nonlinear parameter from $f_{NL}=-5/4$ to $f_{NL}=100$ can be obtained. 

We also found that for kinematically blocked curvaton, the coupling constant can be as large as ${\cal O}(1)$. This may make curvaton model building easier to connect with particle physics. However, for such a large coupling, we may have to consider quantum corrections or thermal effects to the potential \cite{Enqvist:2011jf}. Nevertheless these effects are model dependent and our idea can work at least in models where these effects are not significant. On the other hand, our model suggests that for some fields which conventionally may be regarded not as a curvaton may still be able to produce non-negligible primordial curvature perturbations.

\label{conclu}

\acknowledgments
CML was supported by the NSC under grant No. NSC 99-2811-M-007-068 and
CSC was supported by the National Center of Theoretical Sciences of Taiwan (NCTS).

\end{document}